\documentclass[english,american]{revtex4-2}
\usepackage[T1]{fontenc}
\usepackage[latin9]{inputenc}
\usepackage{babel}
\usepackage{float}
\usepackage{amsmath}
\usepackage{amsthm}
\usepackage{amssymb}
\usepackage{graphicx}
\usepackage{esint}
\usepackage[pdfusetitle,
 bookmarks=true,bookmarksnumbered=false,bookmarksopen=false,
 breaklinks=false,pdfborder={0 0 1},backref=false,colorlinks=false]
 {hyperref}
\hypersetup{
 colorlinks,linkcolor=blue,citecolor=blue,urlcolor=blue}

\makeatletter

\newcommand*\LyXZeroWidthSpace{\hspace{0pt}}
\providecommand{\tabularnewline}{\\}

\makeatother

\begin{document}
\selectlanguage{english}%
\global\long\def\ket#1{\left|#1\right\rangle }%

\global\long\def\bra#1{\left\langle #1\right|}%

\global\long\def\ketL#1{\left.\left|#1\right\rangle \right\rangle }%

\global\long\def\braL#1{\left\langle \left\langle #1\right|\right.}%

\global\long\def\braket#1#2{\left\langle #1\left|#2\right.\right\rangle }%

\global\long\def\ketbra#1#2{\left|#1\right\rangle \left\langle #2\right|}%

\global\long\def\braOket#1#2#3{\left\langle #1\left|#2\right|#3\right\rangle }%

\global\long\def\mc#1{\mathcal{#1}}%

\global\long\def\nrm#1{\left\Vert #1\right\Vert }%

\title{Over-rotation coherent error induced by pseudo-twirling of quantum gates}
\author{Tanmoy Pandit}
\email{tanmoy.pandit@mail.huji.ac.il }
\affiliation{Fritz Haber Research Center for Molecular Dynamics, Hebrew University
of Jerusalem, Jerusalem 9190401, Israel}
\author{Raam Uzdin}
\email{raam@mail.huji.ac.il}
\affiliation{Fritz Haber Research Center for Molecular Dynamics, Hebrew University
of Jerusalem, Jerusalem 9190401, Israel}
\begin{abstract}
{Quantum error mitigation schemes (QEM) have greatly enhanced the performance of quantum computers, mostly by reducing errors caused by interactions with the environment. Nevertheless, the presence of coherence errors, typically arising from miscalibration and inter-qubit crosstalk, is a significant challenge to the scalability of quantum computing. Such errors are often addressed using a refined Pauli twirling scheme called Randomized Compiling (RC) that converts the coherent errors into incoherent errors that can then be mitigated by conventional QEM. Unfortunately for multi-qubit gates, RC is restricted to Clifford gates such as CNOT and CPHASE. However, it has been demonstrated experimentally that a direct implementation of multi-qubit non-Clifford gates, i.e. without using multi-qubit Clifford gates, has reduced the depth of the circuit by a factor of four and more. Recently, a framework called pseudo-twirling (PST) for treating coherent error in multi-qubit non-Clifford gates has been introduced and experimentally demonstrated. We show analytically that a higher order correction to the existing PST theory yields an over-rotation coherent error generated by the PST protocol itself. This PST effect has no analogue in RC. Although the small induced over-rotation can amount to a significant coherent error in deep circuits, we explain why it does not degrade the performance of the gate. Interestingly, we find that a simplified twirling scheme that was introduced and exploited experimentally by Kim et al. also displays an induced over-rotation. We study the conditions under which the two twirling schemes display the same over-rotation behavior.}
\end{abstract}
\maketitle

\section{Introduction}

Two types of errors impede the performance of quantum computers: 1)
incoherent errors (``noise'') that arise due to the interaction
of the qubits with the environment, and 2) coherent errors that typically
occur due to miscalibration or crosstalk interaction between the qubits.
While incoherent errors can be addressed by a plethora of quantum
error mitigation (QEM) techniques \cite{cai2022quantum,endo2018practical,suzuki2022quantum,temme2017error,Huggins2021,Koczor2021,PhysRevA.102.012426,strikis2021learning,li2017efficient,TEMqem},
until recently, the only proven tools for addressing coherent errors
was Pauli twirling (PT), and its extended version known as 'randomized
compiling' (RC) \cite{wallman2016noise,PhysRevX.11.041039,RCqutrits,cai2019SmallerTwirl}.
Unlike PT, under some restrictions, RC can also handle single-qubit
non-Clifford gates. In PT/RC, an ensemble of random circuits is created
by padding the multi-qubit Clifford gate by Pauli gates, which, in
the absence of coherent error, do not alter the functionality of the
Clifford gate. The ensemble average cancels the odd orders of the
coherent error, and the remaining even orders manifest as incoherent
error that can be addressed by QEM. This effect is expected since
the ensemble average is a mixture of unitaries (if there are no incoherent
errors), thereby reducing the purity of the final state. 

Non-Clifford gates are crucial for achieving quantum advantage because
without them circuits can be efficiently simulated on a classical
computer. In the standard CNOT (or CPhase) paradigm, all non-Clifford
gates are low-noise single-qubit gates. While this choice is sensible,
it has the drawback that any departure from a two-qubit Clifford gate
requires at least two CNOT operations, which substantially increase
the runtime and the exposure to incoherent error mechanisms. Non-Clifford
two-qubit gates naturally arise in lattice simulations (e.g., transverse
Ising model), quantum Fourier transform, QAOA circuits, and more.
In a previous studies, it was experimentally demonstrated that using
multi-qubit non-Clifford (MQNC) gates reduced circuit depth by a factor
of four \cite{kim2023scalable} and six \cite{ScaledZXibm}, leading
to a significant decrease in overall noise levels within the circuit.
Unfortunately, since PT and RC cannot be applied to these gates, they
are susceptible to calibration and crosstalk errors. A formalism called
pseudo twirling (PST) for addressing coherent errors in MQNC gates
was recently introduced \cite{santos2024pseudo}. Earlier descriptions
of the PST protocol, lacking supporting theory (see the discussion
in \cite{santos2024pseudo}), are found in
\cite{layden2023MarkovMontePST,chen2022SimPSTwrong}. 

Another method for addressing coherent and incoherent errors in non-Clifford
gates has been introduced in \cite{laydenPECnonCliff}. Unfortunately,
the sampling overhead of this method is highly non-scalable. Furthermore,
since it is based on noise learning, it is inherently sensitive to
time variations in the noise profile. While PT and PST require the
implementation of many circuits, they do not involve a sampling overhead.
The total number of shots needed to achieve the target accuracy without
twirling is evenly distributed among the different twirling circuits. 

The key differences between PST and PT are: the specific types of
coherent errors addressed, the nature of the resulting noise channel,
and the level of control necessary to execute these protocols. Due
to these distinctions, one method cannot be considered as a special
case of the other. Beginning with the first difference, PT treats
all types of coherent errors by converting them into noise, whereas
PST excludes a particular type of coherent error: controlled mis-rotations.
This difference has an important operational implication. Unlike PT,
PST can be applied during the gate calibration stage. The advantage
of employing PST in high-accuracy calibration protocols was experimentally
demonstrated in \cite{santos2024pseudo}. Secondly, PT transforms
coherent errors into Pauli errors, i.e. the noise has a diagonal form
in the Pauli basis, whereas PST, in the first order, renders the error
channel Hermitian with off-diagonal elements in the Pauli basis. Finally,
PT only requires the ability to implement single-qubit Pauli gates,
whereas PST also requires altering the sign of the driving amplitude.
Yet this operational advantage of PT is irrelevant when it comes to
multi-qubit non-Clifford gates where PT is not applicable.

Our work highlights an intriguing aspect that further distinguishes
PST from PT: the contribution of a second-order Magnus term $(\Omega_{2})$
introduces a small coherent error by causing a slight over-rotation
in the driving field. While this might initially appear problematic,
given that the primary aim of PST is to eliminate coherent errors,
we explain why this specific type of coherent error can generally
be disregarded in most scenarios. The analysis presented in \cite{santos2024pseudo}
is grounded in the first-order Magnus expansion. This approximation
proves effective even in situations where errors are significant \cite{KIKarxiv}.
However, it is possible that for certain purposes, such as high-accuracy
calibration, the influence of the second order term cannot be overlooked.
The objective of this paper is to explicitly determine the impact
of the next order. In doing so, we also establish a clear operational
regime for the first-order approximation. 

\textcolor{black}{The paper starts in Sec. II with some preliminaries and a short review
of the PST formalism. In Sec. III we evaluate the $\Omega_{2}$ term
for arbitrary coherent error. Section IV studies the impact of incoherent
noise and higher order correction using parity arguments. Section V discusses another variant of twirling scheme, \textcolor{black}{we term half-twirling (HT) that was introduced in \cite{kim2023scalable}. We find that this protocol also induces an over-rotation but it may not be captured by the PST over-rotation formula. We give a condition for the PST formula to be valid for the HT protocol.}
Finally,
in Section VI, we conclude and discuss the implications of the $\Omega_{2}$ and HT protocol for practical applications.}

\section{Preliminaries}

\subsection{Introduction To Liouville space}

The evolution of an isolate quantum system is described by the Liouville
von Neumann equation
\begin{equation}
\frac{d}{dt}\rho=-\frac{i}{\hbar}[H,\rho],\label{eq:=000020SE}
\end{equation}
where $\rho$, the density matrix that describes the quantum state,
is a $n\times n$ matrix and $n$ is the Hilbert space dimension of
the system. Liouville space is an alternative formulation where the
density matrix $\rho$ is flattened into a density vector $\ket{\rho}$
of dimension $n^{2}\times1$ \cite{gyamfi2020fundamentals}. In Liouville
space, the quantum dynamics (\ref{eq:=000020SE}) translates into
a Schrödinger-like equation
\begin{equation}
\frac{d}{dt}\ket{\rho(t)}=-\frac{i}{\hbar}\mathcal{H}(t)\ket{\rho(t)},\label{eq_of_motion}
\end{equation}
where the Liouville Hamiltonian $\mathcal{H}(t)$ is related to the
Hilbert space Hamiltonian $H(t)$ through $\mathcal{H}(t)=H(t)\otimes I-I\otimes H(t)^{T}$,
where $\intercal$ stands for transposition. The evolution operator
$\mathcal{U}(t)$ that propagates the quantum state in Liouville space
$\ket{\rho(t)}=\mathcal{H}(t)\ket{\rho(0)}$ can be expressed as 
\begin{equation}
\mathcal{U}(t)=U(t)\otimes U(t)^{*},\label{Ulio}
\end{equation}
where, $U(t)$ is the evolution operator in Hilbert space $\rho(t)=U(t)\rho(0)U(t)^{\dagger}$.
Pauli matrices in Liouville space can either appear as Hamiltonians
or as unitaries, but unlike in Hilbert space, these two forms differ.
Let us denote by $P_{\alpha}$ the tensor product of single-qubit
Pauli matrices $\sigma_{i}$ such that $P_{\alpha}\in\{\sigma_{k}\otimes\sigma_{l}\otimes\sigma_{m}...\}_{k,l,m..\in\{0,x,y,z\}}$.
A Pauli Hamiltonian in Liouville space is given by $\mathcal{H}_{\alpha}=P_{\alpha}\otimes I-I\otimes P_{\alpha}^{\intercal}$,
while according to Eq. (\ref{Ulio}), a Pauli evolution operator has
the form $\mathcal{P}_{\alpha}=P_{\alpha}\otimes P_{\alpha}^{*}$
in Liouville space. These two Pauli forms are related via $\mathcal{P}_{\alpha}=exp(-i\frac{\pi}{2}\mathcal{H}_{\alpha})$.
In particular, we shall use the fact that Pauli unitaries always commute
in Liouville space while two Pauli Hamiltonians in Liouville commute
(anti commute) if their corresponding Paulis in Hilbert space commute
(anti commute). The same holds the commutation relation between Pauli
unitaries $\mathcal{P}_{\alpha}$ and Pauli Hamiltonians $\mathcal{H}_{\alpha}$.
Finally, while $\mathcal{P}_{\alpha}^{2}$ is equal to the identity
operator $\mathcal{H}_{\alpha}^{2}$ is not equal to the identity.

\subsection{Overview of the PST formalism}

\begin{figure}[ht]
\centering \includegraphics[width=0.9\textwidth]{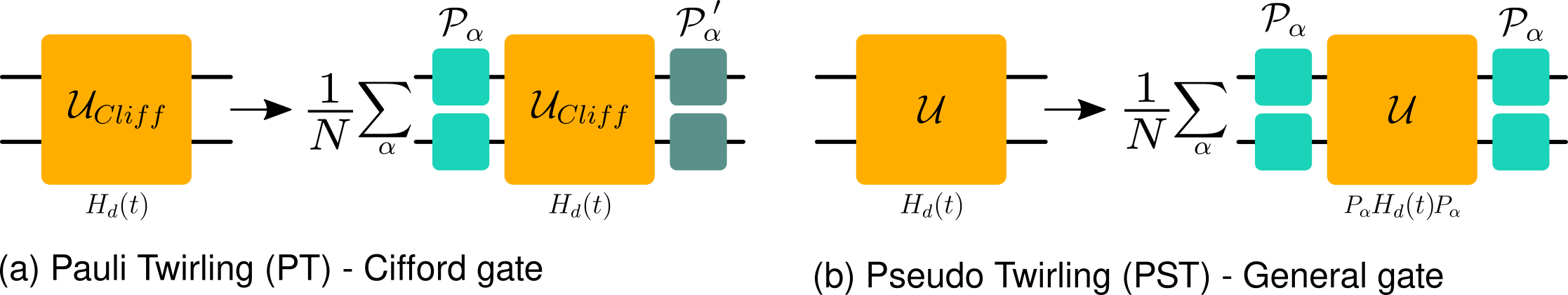} \caption{ (a) Pauli Twirling (PT) and randomized compiling (RC), involve replacing
the original multi-qubit Clifford gate (such as a CNOT) with a collection
of Clifford gates that have been twirled by Pauli gates. These methods
are not suitable to multi-qubit non-Clifford gates. (b) The recently
proposed Pseudo Twirling (PST) scheme effectively reduces coherent
errors in a general gate. This is achieved by applying the same Pauli's
before and after the gate and altering the signs of certain pulses
in the driving field $H_{d}(t)$ that generate the gate based on the
rule $H_{d}(t)\to P_{\alpha}H_{d}(t)P_{\alpha}$. N is the number of different twirling realizations. }\label{fig:illust}
\end{figure}

To illustrate the pseudo-twirling protocol studied in \cite{santos2024pseudo},
we begin with a straightforward example. Consider the following non-Clifford
two-qubit gate $\mathcal{H}_{zz}(\theta)=e^{-i\theta\mathcal{H}_{zz}},\theta\neq k\pi/2$
where $\mathcal{H}_{zz}=P_{zz}\otimes I-I\otimes P_{zz}^{\intercal}$
and $P_{zz}=\sigma_{z}\otimes\sigma_{z}$. In analogy to PT, our goal
is to create an operation that retains the functionality of the ideal
gate. When twirling using $\mathcal{P}_{xx}=P_{xx}\otimes P_{xx}^{*}$,
we find that $\mathcal{P}_{xx}e^{-i\theta\mathcal{H}_{zz}}\mathcal{P}_{xx}=e^{-i\theta\mathcal{H}_{zz}}$
because $\mathcal{P}_{xx}$ commutes with $\mathcal{H}_{zz}$. In
contrast, $\mathcal{P}_{xz}$ anti-commutes with $\mathcal{H}_{zz}$,
resulting in $\mathcal{P}_{xz}e^{-i\theta\mathcal{H}_{zz}}\mathcal{P}_{xz}=e^{+i\theta\mathcal{H}_{zz}}=\mathcal{U}_{zz}(-\theta)$.
The PST protocol leverages the fact that if Pauli operators anti-commute
with the Hamiltonian that generates $U$ (or equivalently $\mathcal{U}$
in Liouville space), the desired transformation can be achieved by
reversing the angle, i.e., changing the sign of the driving fields.
The difference between Pauli twirling and pseudo twirling is shown
in Fig. \ref{fig:illust}.

In a broader context, when the driving Hamiltonian is composed of
multiple Pauli terms, the evolution operator becomes $\mathcal{U}=e^{-i\sum h_{\beta}\mathcal{H}_{\beta}}$.
The Pseudo Twirling (PST) protocol can be mathematically expressed
as 
\begin{equation}
\mathcal{K}_{PST}=\frac{1}{2^{2n}}\sum_{\alpha=1}^{2^{2n}}\mathcal{P}_{\alpha}e^{-i\mathcal{P}_{\alpha}(\sum_{\beta}h_{\beta}\mathcal{H}_{\beta})\mathcal{P}_{\alpha}}\mathcal{P}_{\alpha}=\frac{1}{2^{2n}}\sum_{\alpha=1}^{2^{2n}}\mathcal{P}_{\alpha}e^{-i\sum_{\beta}\text{sgn}(\alpha,\beta)h_{\beta}\mathcal{H}_{\beta}}\mathcal{P}_{\alpha},\label{eq:=000020PST}
\end{equation}
where $\text{sgn}(\alpha,\beta)=\frac{\text{tr}(P_{\alpha}P_{\beta}P_{\alpha}P_{\beta})}{2^{n}}$
equals $\pm1$ if $P_{\alpha}$ and $P_{\beta}$ commute or anti-commute,
respectively. Note that in the definition of $\text{sgn}(\alpha,\beta)$,
Pauli matrices in Hilbert space are used. Mathematically, expression
(\ref{eq:=000020PST}) is straightforward in the absence of coherent
error. Operationally, at the beginning of the protocol, the appropriate
$\mathcal{P}_{\alpha}$ is applied. Next, a modified unitary function
$\mathcal{U}_{\alpha}=e^{-i\sum_{\beta}\text{sgn}(\alpha,\beta)h_{\beta}\mathcal{H}_{\beta}}$
is executed by changing the sign of specific control signals. Finally,
another $\mathcal{P}_{\alpha}$ is applied at the end of the gate.
Importantly, the Hamiltonian of the modified unitary retains the same
terms as the original Hamiltonian, with only the signs of some terms
changed. Thus, under the assumption that the signs of the driving
Hamiltonian terms are controllable, the PST implementation is no more
challenging than implementing the original unitary. 

PST has two distinctive features compared to randomized compiling.
Firstly, it is applicable to non-Clifford gates. Secondly, there is
one specific coherent error it cannot average out: a controlled mis-rotation.
This latter feature is actually useful for calibration purposes (see
\cite{santos2024pseudo}). 

\subsection{PST Theory from Magnus Expansion}

In this work we consider a Hamiltonians of the form
\begin{equation}
\mathcal{H}=\mathcal{H}_{\beta}+\xi\mathcal{H}_{coh},
\end{equation}
where $\mathcal{H}_{\beta}$ is the drive Hamiltonian that in the
absences of coherent error generates the desired unitary $\mathcal{U}=e^{-i\tau\mathcal{H}_{\beta}}$.
$\mathcal{H}_{coh}=\sum_{\gamma}h_{\gamma}\mathcal{H}_{\gamma}$ represents
the coherent errors we wish to mitigate. Using the Magnus expansion
\cite{blanes2009magnus} in the interaction picture, one can write
the evolution operator at time $\tau$ as
\begin{align}
\mathcal{U}_{\xi}(\tau) & =\mathcal{U}(\tau)e^{\sum_{n=1}^{\infty}\Omega_{n}},
\end{align}
where $\tau$ is the duration of the driving pulse, $\mathcal{U}(t)=\exp(-i\mathcal{H_{\beta}}t)$
, 
\begin{align}
\Omega_{1} & =-i\intop_{0}^{\tau}\mathcal{H}_{int}(t)dt,\\
\Omega_{2} & =\frac{(-i)^{2}}{2}\int_{0}^{\tau}dt_{1}\int_{0}^{t_{1}}dt_{2}[\mathcal{H}_{int}(t_{1}),\mathcal{H}_{int}(t_{2})],
\end{align}
and $\mathcal{H}_{int}(t)=\mathcal{U}(t)^{\dagger}\mathcal{H}_{coh}\mathcal{U}(t)$
is the coherent error in the interaction picture. Since $\Omega_{n}$
is proportional to $\xi^{n}$ and $\xi\ll1$ when the coherent error
in each gate is small, in previous analysis of the PST protocol all
term but the first Magnus terms were neglected. In general, the higher
order Magnus terms are difficult to calculate and have a small impact
when $\xi\ll1$. However here we show that the second order Magnus
term leads to a surprisingly simple result which is the focus of the
present paper. 

By Taylor expansion of the Magnus term or directly from the Dyson
series, one gets that a expansion of order $\xi^{2}$ of $\mathcal{U}_{\xi}(\tau)$
reads
\begin{align}
\mathcal{U}_{\xi}(\tau) & =\mathcal{U}(\tau)(I+\Omega_{1}+\Omega_{1}^{2}/2+\Omega_{2})+O(\xi^{3}).
\end{align}

The application of pseudo twirling leads to following evolution operator
\cite{santos2024pseudo}.
\begin{equation}
\mathcal{K}_{pst}=\mathcal{U}_{0}\frac{1}{2^{2n}}\sum_{\alpha}[I+\Omega_{1,\alpha}+\frac{1}{2}\Omega_{1,\alpha}^{2}+\Omega_{2,\alpha}]+O(\xi^{3}),
\end{equation}
where
\begin{equation}
\Omega_{1,\alpha}=\int_{0}^{\tau}\mathcal{U}(t)^{\dagger}\mathcal{P}_{\alpha}\mathcal{H}_{coh}\mathcal{P}_{\alpha}\mathcal{U}(t)dt,
\end{equation}
and 
\begin{equation}
\Omega_{2,\alpha}(\mathcal{H}_{coh})=\frac{(-i)^{2}}{2}\int_{0}^{T}dt_{1}\int_{0}^{t_{1}}dt_{2}[\mathcal{U}(t_{1})^{\dagger}\mathcal{P}_{\alpha}\mathcal{H}_{coh}\mathcal{P}_{\alpha}\mathcal{U}(t_{1}),\mathcal{U}(t_{2})^{\dagger}\mathcal{P}_{\alpha}\mathcal{H}_{coh}\mathcal{P}_{\alpha}\mathcal{U}(t_{2})].\label{eq:=000020Omega2=000020alpha}
\end{equation}
In \cite{santos2024pseudo} it was shown that PST averages to zero
the $\Omega_{1,\alpha}$ term, and the averaging of $\Omega_{1,\alpha}^{2}$
over $\alpha$ leads to Lindblad dissipation terms that represent
the noise associated with PST, i.e. the conversion of coherent error
into incoherent errors. Next, we turn to our attention to the averaging
of $\Omega_{2,\alpha}$. 

\section{The PST induced over-rotation}

Using $\mc H_{coh}=\sum_{\gamma}h_{\gamma}\mc H_{\gamma}$, we get
\begin{align}
\frac{1}{2^{2n}}\sum_{\alpha}\Omega_{2,\alpha}=\frac{(-i)^{2}}{2}\frac{1}{2^{2n}}\sum_{\alpha}\int_{0}^{T}dt_{1}\int_{0}^{t_{1}}dt_{2}[\mathcal{U}(t_{1})^{\dagger}\mathcal{P}_{\alpha}\sum_{\gamma}h_{\gamma}\mc H_{\gamma}\mathcal{P}_{\alpha}\mathcal{U}(t_{1}),\mathcal{U}(t_{2})^{\dagger}\mathcal{P}_{\alpha}\sum_{\gamma'}h_{\gamma'}\mc H_{\gamma'}\mathcal{P}_{\alpha}\mathcal{U}(t_{2})]\nonumber \\
=\frac{(-i)^{2}}{2}\sum_{\gamma}h_{\gamma}^{2}\int_{0}^{T}dt_{1}\int_{0}^{t_{1}}dt_{2}[\mathcal{U}(t_{1})^{\dagger}\mc H_{\gamma}\mathcal{U}(t_{1}),\mathcal{U}(t_{2})^{\dagger}\mc H_{\gamma}\mathcal{U}(t_{2})]\nonumber \\
+\frac{(-i)^{2}}{2}\frac{1}{2^{2n}}\sum_{\gamma\neq\gamma'}\sum_{\alpha}sgn(\alpha,\gamma)sgn(\alpha,\gamma')\int_{0}^{T}dt_{1}\int_{0}^{t_{1}}dt_{2}[\mathcal{U}(t_{1})^{\dagger}h_{\gamma}\mc H_{\gamma}\mathcal{U}(t_{1}),\mathcal{U}(t_{2})^{\dagger}h_{\gamma'}\mc H_{\gamma'}\mathcal{U}(t_{2})]. \label{omega2cross}
\end{align}
However, since $\sum_{\alpha}sgn(\alpha,\gamma)sgn(\alpha,\gamma')=0$
as shown in the Appendix of \cite{santos2024pseudo}, the last term
is equal to zero and we finally obtain
\begin{equation}
\frac{1}{2^{2n}}\sum_{\alpha}\Omega_{2,\alpha}=\sum_{\gamma}h_{\gamma^{2}}\Omega_{2}(\mathcal{H}_{\gamma}),
\end{equation}
where
\begin{equation}
\Omega_{2}(\mathcal{H}_{\gamma})=-\frac{1}{2}\int_{0}^{\tau}dt_{1}\int_{0}^{t_{1}}dt_{2}[\mathcal{U}(t_{1})^{\dagger}\mathcal{H}_{\gamma}\mathcal{U}(t_{1}),\mathcal{U}(t_{2})^{\dagger}\mathcal{H}_{\gamma}\mathcal{U}(t_{2})].\label{eq:=000020Omega2=000020Hgamma}
\end{equation}
That is, since the cross terms vanish, it is possible to consider
the contribution of each $\mathcal{H}_{\gamma}$ separately and add
them with $h_{\gamma^2}$ weights. However, if a certain coherent
error element $\mathcal{H}_{\gamma'}$ commutes with the driving Pauli
$[H_{\gamma'},H_{\beta}]=0$, it holds that $\mathcal{U}(t)^{\dagger}\mathcal{H}_{\gamma'}\mathcal{U}(t)=\mathcal{H}_{\gamma'}$
and therefore $\Omega_{2}(\mathcal{H}_{\gamma'})=0$. Thus, these
elements can be omitted from the summation and we have 
\begin{equation}
\frac{1}{2^{2n}}\sum_{\alpha}\Omega_{2,\alpha}=\sum_{\gamma\in\{\gamma^{+}\}}h_{\gamma^{2}}\Omega_{2}(\mathcal{H}_{\gamma}),
\end{equation}
where $\{\gamma^{+}\}$ is the set of $P_{\gamma}$ that anti-commutes
with $P_{\beta}$. To simplify the integrand in (\ref{eq:=000020Omega2=000020Hgamma})
we start with the dressed Hamiltonian expression
\begin{align}
\mathcal{U}(t)^{\dagger}\mathcal{H}_{\gamma}\mathcal{U}(t) &= U(-t) \otimes U(-t)^{*} (P_{\gamma} \otimes I - I \otimes P_{\gamma}^{*}) U(t) \otimes U(t)^{*} \nonumber \\
&= U(-t) P_{\gamma} U(t) \otimes I - I \otimes U(-t)^{*} P_{\gamma}^{*} U(t)^{*}.
\end{align}

Using $U(t)=\exp(-itP_{\beta})$ and $\{P_{\gamma},P_{\beta}\}=0$
(since $\gamma\in\{\gamma^{+}\}$) we get the relation 
\begin{equation}
P_{\gamma}U(t)=P_{\gamma}U(t)P_{\gamma}P_{\gamma}=U(-t)P_{\gamma},
\end{equation}
and consequently
\begin{equation}
\mathcal{U}(t)^{\dagger}\mathcal{H}_{\gamma}\mathcal{U}(t)=U(-2t)P_{\gamma}\otimes I-I\otimes U(-2t)^{*}P_{\gamma}^{*}.
\end{equation}
Using this result in the integrand of Eq. (\ref{eq:=000020Omega2=000020Hgamma})
leads to
\begin{align}
[\mathcal{U}(t_{1})^{\dagger}\mathcal{H}_{\gamma}\mathcal{U}(t_{1}),\mathcal{U}(t_{2})^{\dagger}\mathcal{H}_{\gamma}\mathcal{U}(t_{2})]= & [U(-2t_{1})P_{\gamma},U(-2t_{2})P_{\gamma}]\otimes I\nonumber \\
+ & I\otimes[U(-2t_{1})^{*}P_{\gamma}^{*},U(-2t_{2})^{*}P_{\gamma}^{*}].
\end{align}
For the first term we obtain
\begin{equation}
[U(-2t_{1})P_{\gamma},U(-2t_{2})P_{\gamma}]=U(-2t_{1})P_{\gamma}U(-2t_{2})P_{\gamma}-U(-2t_{2})P_{\gamma}U(-2t_{1})P_{\gamma}=2i\sin[2(t_{2}-t_{1})]P_{\beta},
\end{equation}
and a similar expression holds for the second term. Consequently we
get

\begin{align}
[\mathcal{U}(t_{1})^{\dagger}\mathcal{H}_{\gamma}\mathcal{U}(t_{1}),\mathcal{U}(t_{2})^{\dagger}\mathcal{H}_{\gamma}\mathcal{U}(t_{2})] & =2i\sin[2(t_{2}-t_{1})]P_{\beta}\otimes I-2iI\otimes\sin[2(t_{2}-t_{1})]P_{\beta}^{*}\nonumber \\
 & =-2i\sin[2(t_{2}-t_{1})]\mathcal{H}_{\beta}.
\end{align}
Finally, after the double integration, and using $\text{sinc(x)}=\frac{\sin x}{x}$
we obtain our main result 
\begin{align}
\frac{1}{2^{2n}}\sum_{\alpha}\Omega_{2,\alpha} & =-i\frac{2\tau-\sin(2\tau)}{4}(\sum_{\gamma\in\{\gamma^{+}\}}h_{\gamma}^{2})\mathcal{H}_{\beta},\nonumber \\
 & =-i\tau\frac{1-\text{sinc}(2\tau)}{2}(\sum_{\gamma\in\{\gamma^{+}\}}h_{\gamma}^{2})\mathcal{H}_{\beta},
\end{align}
and therefore the effective drive Hamiltonian is 
\begin{equation}
\mathcal{H}_{eff}(\tau)=[1+\frac{1-\text{sinc}(2\tau)}{2}(\sum_{\gamma\in\{\gamma^{+}\}}h_{\gamma}^{2})]\mathcal{H}_{\beta}.\label{eq:=000020main=000020result}
\end{equation}
Since $\frac{1-\text{sinc}(2\tau)}{2}\ge0$ for any $\tau$, the term
$[1+\frac{1-\text{sinc}(2\tau)}{2}(\sum_{\gamma\in\{\gamma^{+}\}}h_{\gamma}^{2})]$
can be considered as an amplitude amplification factor, or alternatively,
as a relative over-rotation. At first, it may seem quite striking
that after PST, all the various coherent errors manifest simply as
modification of the coefficient in front of the driving Hamiltonian.
However, we argue that this must be the case. Had a different term
survived after PST we could have applied PST again to eliminate it.
Note that even if the survived term had the same sign dependence as
that of the drive, PST would still eliminated it (see Sec. III.A in
\cite{santos2024pseudo}). However, PST of a PST is just a regular
PST. This leads to the conclusion that when it comes to residual coherent errors only mis-rotation of the drive  is possible.
Although we do not study it here, we expect the same logic to be valid
for $\Omega_{3}$ and higher orders in the Magnus expansion.

To verify our theoretical result, Eq.~\eqref{eq:=000020main=000020result},
we consider a non-Clifford ZX gate with uncontrolled coherent errors,
i.e. independent of the driving $\tau$. The Hamiltonian 
\begin{align}
H & =\sigma_{z}\otimes\sigma_{x}+(\xi_{xx}\sigma_{x}\otimes\sigma_{x}+\xi_{yy}\sigma_{y}\otimes\sigma_{y}+\xi_{zz}\sigma_{z}\otimes\sigma_{z}+\xi_{xz}\sigma_{x}\otimes\sigma_{z}).
\end{align}
is propagated for a duration $\tau$. The resulting evolution operator
is
\begin{equation}
\mathcal{K}_{pst}=e^{-i\tau\mathcal{H}_{eff}+\mathcal{L}_{eff}}.
\end{equation}
Next, we calculate the weight of different terms in the Hamiltonian
(effective Hamiltonian) before (after) performing PST. We choose the
coherent error in such a way that three components (XX, ZZ, and YX)
anti-commute with the original Hamiltonian and one term (YY) commutes
with the driving Hamiltonian. Table \ref{tab:=000020numerics} validates
the theoretical prediction Eq.~\eqref{eq:=000020main=000020result}
of the PST induced over-rotation. The simulation parameters are $\xi_{xx}=0.2$,
$\xi_{yy}=0.6$, $\xi_{zz}=0.2$, $\xi_{yx}=0.4$ and $\tau=0.5$.

\begin{table}[htbp]
\centering %
\begin{tabular}{|c|c|c|c|c|c|c|}
\hline 
 & XX & YY & ZZ & YX & ZX (numerics) & ZX PST (theoretical)\tabularnewline
\hline 
No PST & 0.2 & 0.6 & 0.2 & 0.4 & 1 & -\tabularnewline
\hline 
PST & 0 & 0 & 0 & 0 & 1.0207  & 1.019\tabularnewline
\hline 
\end{tabular}\caption{Comparison of the effective Hamiltonian terms in the evolution operators
with and without PST for various coherent error components. While
the PST completely eliminates the coherent error in XX,YY,ZZ and YX,
it generates an effective increase in the amplitude of the ZX drive
from $1$ to $1.0207$ ($2\%$ deviation). The analytical formula
(\ref{eq:=000020main=000020result}) yields the value $1.019$ which
matches the numerical value with $99.83\%$ accuracy. Consistently
with (\ref{eq:=000020main=000020result}), the YY term does not contribute
to the over-rotation since it commutes with the ZX drive.}\label{tab:=000020numerics}
\end{table}

\section{Parity and higher orders}

In this section, we employ a parity argument in order to study third
order effects without explicit calculations. In the original frame,
i.e. before the transition to the interaction frame, the PST evolution
operator is 
\begin{equation}
\mathcal{K}_{pst}=\frac{1}{2^{2n}}\sum_{\alpha}\mathcal{P_{\alpha}}e^{-i\tau\mathcal{P_{\alpha}}\mathcal{H}_{\beta}\mathcal{P_{\alpha}}-i\delta\tau\mathcal{H}_{coh}}\mathcal{P_{\alpha}}=\frac{1}{2^{2n}}\sum_{\alpha}e^{-i\tau H_{\beta}-i\delta\tau\mathcal{P_{\alpha}}\mathcal{H}_{coh}\mathcal{P_{\alpha}}}.
\end{equation}
Next, we consider the case where there exist a Pauly matrix $P_{\gamma}$
that anti-commutes with $H_{coh}$. As a result $\mathcal{P_{\gamma}}\mathcal{H}_{coh}\mathcal{P_{\gamma}}=-\mathcal{H}_{coh}$.
For any element $\alpha$ in the set $\{\alpha\}$ we choose another
element $\bar{\alpha}$ defined as $\mathcal{P}_{\bar{\alpha}}=\mathcal{P_{\gamma}}\mathcal{P}_{\alpha}$.
Thus, if we carry out the sum over $\alpha$ in pair of $\alpha$
and $\bar{\alpha}$ we get that each pair form an even function of
$\delta$: 
\begin{align}
\mathcal{K}_{pst}= & \sum_{\alpha}^{2^{2n}/2}[e^{-i\tau H_{\beta}-i\delta\tau\mathcal{P_{\alpha}}\mathcal{H}_{coh}\mathcal{P_{\alpha}}}+e^{-i\tau H_{\beta}-i\delta\tau\mathcal{P_{\bar{\alpha}}}\mathcal{H}_{coh}\mathcal{P_{\bar{\alpha}}}}]\nonumber \\
= & \sum_{\alpha}^{2^{2n}/2}[e^{-i\tau H_{\beta}-i\delta\tau\mathcal{P_{\alpha}}\mathcal{H}_{coh}\mathcal{P_{\alpha}}}+e^{-i\tau H_{\beta}+i\delta\tau\mathcal{P_{\alpha}}\mathcal{H}_{coh}\mathcal{P_{\alpha}}}].
\end{align}
Thus, $\mathcal{K}_{pst}$ has the form of an even function in $\delta$:
$f(x,\delta)+f(x,-\delta)$. Consequently, all odd order term in $\delta$
in the Taylor expansion must be zero. In particular the third order
in $\delta$ has to be zero. 

\begin{figure}[H]
\centering\includegraphics[width=8cm]{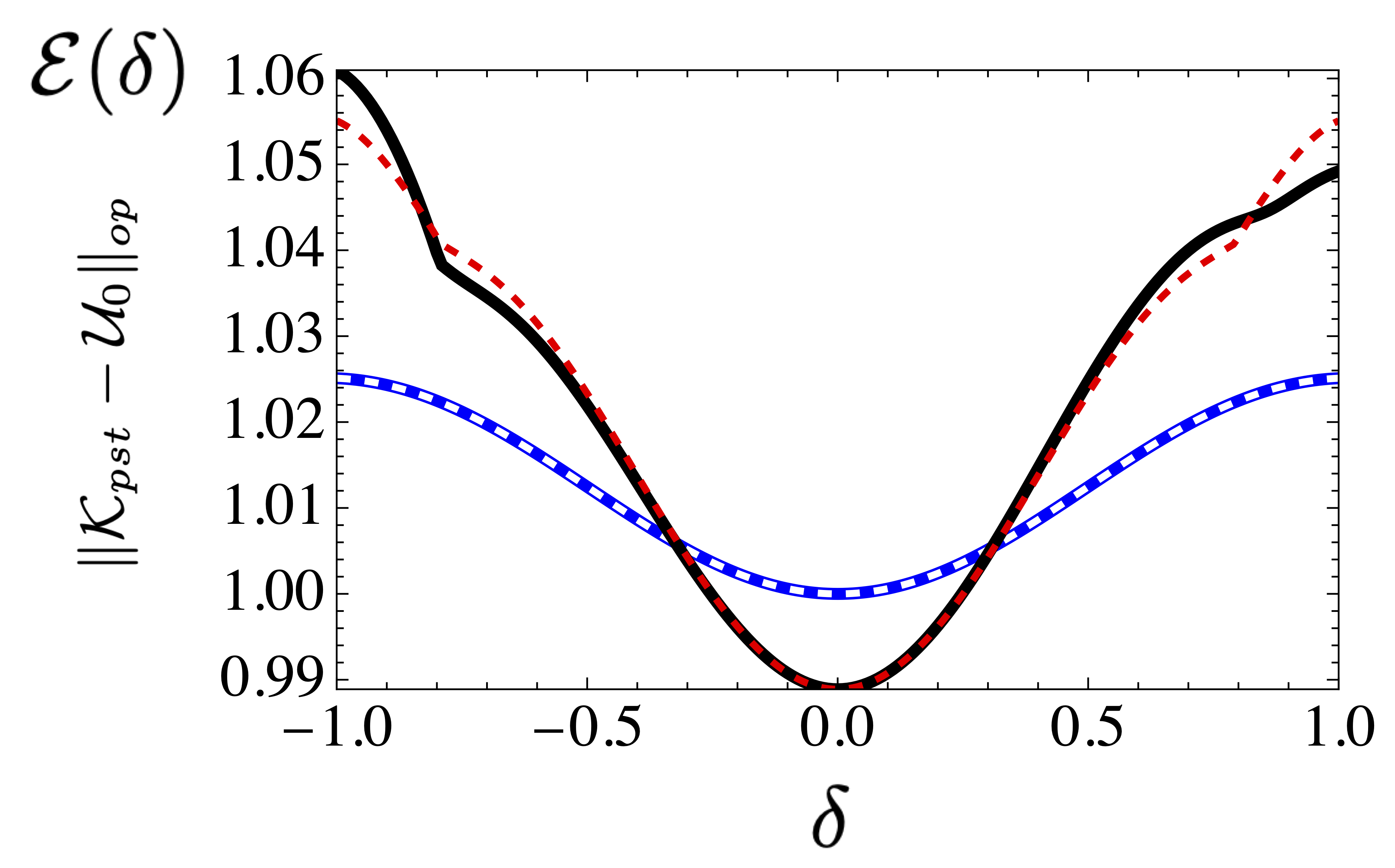}\caption{ Comparison of Pauli Z error (blue) and amplitude damping error (black)
noise models as a function of the strength of the coherent error $\delta$.
The $y$ axis correspond to operator norm of the deviation of the
PST channel $\mathcal{K}_{pst}$ from the ideal channel $\mathcal{U}_{0}$
$\mathcal{E}(\delta)=\|\mathcal{K}_{pst}-\mathcal{U}_{0}\|_{op}$.
The dashed lines represent a symmetrized reference. As explained in
the main text, non-Pauli channel can break the $\mathcal{E}(\delta)=\mathcal{E}(-\delta)$
symmetry, leading to small yet non trivial contribution from expansion
terms of order $\delta^{2n+1}$. }\label{fig:errors}
\end{figure}

Importantly, this symmetry can be broken by the present of noise:
\[
\mathcal{K}_{pst}=\sum_{\alpha}^{2^{2n}/2}e^{-i\tau H_{\beta}-i\delta\tau\mathcal{P_{\alpha}}\mathcal{H}_{coh}\mathcal{P_{\alpha}}+\zeta\mathcal{P_{\alpha}}\mathcal{L}\mathcal{P_{\alpha}}}+e^{-i\tau H_{\beta}+i\delta\tau\mathcal{P_{\alpha}}\mathcal{H}_{coh}\mathcal{P_{\alpha}}+\mathcal{\zeta P_{\bar{\alpha}}}\mathcal{L}\mathcal{P_{\bar{\alpha}}}}.
\]
Since in general $\mathcal{P_{\alpha}}\mathcal{L}\mathcal{P_{\alpha}}\neq\mathcal{P_{\bar{\alpha}}}\mathcal{L}\mathcal{P_{\bar{\alpha}}}$,
the even parity no longer holds. Interestingly, in the important case
of Pauli noise $\mathcal{P_{\bar{\alpha}}}\mathcal{L}\mathcal{P_{\bar{\alpha}}}=\mathcal{P_{\alpha}}\mathcal{L}\mathcal{P_{\alpha}}=\mathcal{L}$,
and therefore the symmetry still holds, leading to $O(\delta^{3})=0$.
In Fig. \ref{fig:errors} we plot the error $\mathcal{E}(\delta)=\|\mathcal{K}_{pst}-\mathcal{U}_{0}\|_{op}$
of the PST channel with respect to the ideal channel. We used the
same $\mathcal{H}_{coh}$ as in Table \ref{tab:=000020numerics} but
with added coefficient $\delta$ that enables control over the amplitude
of and the sign of the coherent error. The blue line shows that for
a Pauli Z noise (decoherence), $\mathcal{E}(\delta)$ is indistinguishable
from the symmetrized dashed line $\frac{1}{2}[\mathcal{E}(\delta)+\mathcal{E}(-\delta)]$.
When the noise is an amplitude damping channel (black line), the symmetry
of $\mathcal{E}(\delta)$ is broken and $\mathcal{E}(\delta)\neq\mathcal{E}(-\delta)$.
However, the difference is small. To observe this difference we use
$\zeta=3$ which corresponds to an extreme decay. We also used $\tau=5/2$
for making the effect more visible. 

\textcolor{black}{\section{Analysis of the Half-Twirling Protocol}}
In this section, we study a protocol we refer to as "half-twirling" (HT), where only the subset of Pauli operators that commute with the drive are involved in the twirling process. This protocol has been experimentally implemented, as reported in \cite{kim2023scalable}, but without a rigorous analytical justification or a detailed description of the resulting noise channel. Specifically, it is not immediately evident whether the over-rotation effect, discussed in previous sections, \textcolor{black}{ 
exists and whether it retains the same analytical form as in the PST case}. We define $\beta^{\pm}$ as the sets of $\alpha$'s such that $\mathcal{P}_{\alpha}\mathcal{H}_{\beta}\mathcal{P}_{\alpha} = \pm\mathcal{H}_{\beta}$ ($sgn(\alpha, \beta) = \pm1$). \textcolor{black}{In the derivation in Appendix 2 of \cite{santos2024pseudo} it is shown that}
\begin{equation}
\frac{1}{2^{n}} \sum_{\alpha \in \beta^+} \mathcal{P}_\alpha \mathcal{H}_{\text{coh}} \mathcal{P}_\alpha = \frac{1}{2^{n}} \sum_{\alpha \in \beta^-} \mathcal{P}_\alpha \mathcal{H}_{\text{coh}} \mathcal{P}_\alpha = 0,
\end{equation} \textcolor{black}{hence, half-twirling} also suppresses coherent error in the first order. However, note that it cannot  eliminate uncontrolled coherent errors that are proportional to the drive $\mathcal{H}_\beta$.

To analyze the effect of half twirling on the second-order Taylor approximation as discussed in \cite{santos2024pseudo}, we start by noting that \textcolor{black}{the expression}
\begin{equation}
\int_{0}^{T} \mathcal{U}(t)^{\dagger} \mathcal{P}_{\alpha} \mathcal{H}_{\text{coh}} \mathcal{P}_{\alpha} \mathcal{U}(t) \, dt,
\end{equation}

\textcolor{black}{retains the structure of a Hamiltonian in Liouville space, represented as $
H_{eff,\alpha} \otimes I - I \otimes H_{eff,\alpha}^{\intercal},
$}

where,
\begin{equation}
H_{eff,\alpha} = \int_{0}^{T} U(t)^{\dagger} P_{\alpha} H_{\text{coh}} P_{\alpha} U(t) \, dt.
\end{equation}

Utilizing this observation, the expression
\begin{equation}
\left[-i \int_{0}^{T} \mathcal{U}(t)^{\dagger} \mathcal{P}_{\alpha} \mathcal{H}_{\text{coh}} \mathcal{P}_{\alpha} \mathcal{U}(t) \, dt\right]^{2},
\end{equation}transforms into
\begin{equation}
-(H_{eff,\alpha} \otimes I - I \otimes H_{eff,\alpha}^{\intercal})^{2} = 2 H_{eff,\alpha} \otimes H_{eff,\alpha}^{\intercal} - H_{eff,\alpha}^{2} \otimes I - I \otimes (H_{eff,\alpha}^{2})^{\intercal}.
\end{equation}
Given that $H_{eff,\alpha}$ is Hermitian, this \textcolor{black}{exhibits a Lindblad form}:
\begin{equation}
\mathcal{L}(A) = A \otimes A^{*} - \frac{1}{2} (A^{\dagger} A) \otimes I - \frac{1}{2} I \otimes (A^{\dagger} A)^{\intercal},
\end{equation}
where the Hermitian dissipator is $A = H_{eff,\alpha}$.
This statement regarding the Hermitian nature of the noise holds for each $\alpha$. Consequently, restricting $\alpha$ to either $\beta^{+}$ or $\beta^{-}$ does not affect the validity of the Hermitian property of the noise channel, even in the case of half-twirling. Next, we revisit the derivation of the over-rotation effect in the context of half-twirling. In the original derivation, we used the identity $\sum_{\alpha=1}^{2^n} sgn(\alpha,\gamma) sgn(\alpha,\gamma') = 0$ for $\gamma \neq \gamma'$ to assert that the contribution from the cross terms is zero. \textcolor{black}{ 
While this property does not hold for the HT case we find that when the overall coherent error anti-commutes with the driving Hamiltonian, and the different error components of the coherent error commute with each other, the cross terms still vanish.
}

\textcolor{black}{
We begin by examining the cross term of Eq. (\ref{omega2cross}) in the in the integrand }of Eq. (\ref{eq:=000020Omega2=000020Hgamma}):\[
\left[\mathcal{U}(t_{1})^{\dagger} h_{\gamma} \mathcal{H}_{\gamma} \mathcal{U}(t_{1}), \mathcal{U}(t_{2})^{\dagger} h_{\gamma'} \mathcal{H}_{\gamma'} \mathcal{U}(t_{2})\right].
\] \textcolor{black}{Referring to the derivation in Appendix III of ref} \cite{santos2024pseudo}, we can express the first term  and second term of the commutator as:
\begin{align}
    e^{+i t_j \mathcal{H}_\beta} \mathcal{H}_\gamma e^{-i t_j \mathcal{H}_\beta} &= \frac{1+\operatorname{sgn}(\gamma, \beta)}{2} \mathcal{H}_\gamma \nonumber + \frac{1-\operatorname{sgn}(\gamma, \beta)}{2} \left( P_\gamma \otimes I \, e^{-i 2t_j P_\beta \otimes I} \right. \nonumber \\
    &\quad \left. - I \otimes P_\gamma^* \, e^{+i 2t_j I \otimes P_\beta^*} \right).
\end{align}
Where, $j \in{1,2}$. \textcolor{black}{Next, we restrict ourselves in the case where, $\gamma$ and $\beta$ anti-commute with each other, as a result we have \(\operatorname{sgn}(\gamma, \beta) + 1 = 0\). Thus, we can write:}
\begin{equation}
\begin{aligned}
&\left[\mathcal{U}(t_{1})^{\dagger} h_{\gamma} \mathcal{H}_{\gamma} \mathcal{U}(t_{1}), \mathcal{U}(t_{2})^{\dagger} h_{\gamma'} \mathcal{H}_{\gamma'} \mathcal{U}(t_{2})\right] \\
&\quad = h_{\gamma} h_{\gamma'} \bigg( \left[ P_\gamma \otimes I \, e^{-i 2t_1 P_\beta \otimes I} - I \otimes P_\gamma^* \, e^{+i 2t_1 I \otimes P_\beta^*}, \right. \\
&\qquad \left. P_\gamma' \otimes I \, e^{-i 2t_2 P_\beta \otimes I} - I \otimes P_\gamma'^* \, e^{+i 2t_2 I \otimes P_\beta^*} \right] \bigg).
\end{aligned}
\end{equation}

Since there are no terms associated with \(\mathcal{H}_{\gamma,\gamma'}\), and assuming the different components of the coherent error commute among themselves, we have \(\operatorname{sgn}(\gamma, \gamma') = 1\), which implies \(P_\gamma' P_\gamma = P_\gamma P_\gamma'\). Consequently, the commutator corresponding to the cross-term vanishes for each \(\alpha\) value. Thus, we are left with the first term of Equation \ref{omega2cross}, which represents the same over-rotation effect as obtained for the PST. \textcolor{black}{Therefore, the over-rotation effect remains the same in the case when the coherent error anti-commutes with the driving Hamiltonian, and the error components of the coherent error commute among themselves.}

\textcolor{black}{Unlike PST in HT we made the additional assumption that the anti-commuting errors commute among themselves. In the next section we answer whether this is a limitation of our derivation or a needed condition for achieve the PST form obtained before.} \textcolor{black}{To validate our theoretical findings} for  the \textcolor{black}{half-twirling (HT)} case, we examine a non-Clifford ZX gate subject to uncontrolled coherent errors, which are independent of the driving parameter $\tau$. The Hamiltonian is given by
\begin{align}
H &= \sigma_{z} \otimes \sigma_{x} + (\xi_{zz} \sigma_{z} \otimes \sigma_{z} + \xi_{xx} \sigma_{x} \otimes \sigma_{x}) .\label{HamilHT}
\end{align}
This Hamiltonian evolves over a duration $\tau$. \textcolor{black}{The corresponding evolution operator is}
\begin{equation}
\mathcal{K} = e^{-i\tau\mathcal{H}_{eff} + \mathcal{L}_{eff}}.
\end{equation}
Next, we evaluate the contribution of various terms in the Hamiltonian (effective Hamiltonian) both before and after applying HT. We select coherent errors such that three components (ZZ and XX) anti-commute with the original Hamiltonian, and (ZZ and XX commute among themselves). Table \ref{half1} confirms the theoretical prediction Eq.~\eqref{eq:=000020main=000020result} regarding PST-induced over-rotation. The simulation parameters for the Hamiltonian in the Eq. (\ref{HamilHT}) are set to $\xi_{zz} = 0.2$, $\xi_{xx} = 0.2$ and $\tau=0.5$.

\begin{table}[htbp]
\centering
\begin{tabular}{|c|c|c|c|c|c|}
\hline 
 & ZZ & XX & ZX (numerics) & ZX HT (theory)& ZX PST (theory ) \tabularnewline
\hline 
No Twirling& 0.2 & 0.2 & 1.0 & - &-\tabularnewline
\hline 
PST & 0 & 0 & 1.02332 & -&1.02181\tabularnewline
\hline 
HT & 0 & 0 & 1.02332 &1.02181&-
\tabularnewline
\hline 
 
\end{tabular}

\caption{Comparison of the effective Hamiltonian terms in the evolution operators with and without PST and Half-Twirling (HT) for various coherent error components. While Half-Twirling  (HT) and PST completely eliminates the coherent errors in ZZ and XX, it induces a slight increase in the amplitude of the ZX drive from 1 to 1.02332 (2.33\% deviation). The analytical formula (\ref{eq:=000020main=000020result}) gives the value 1.02181, which matches the numerical result with 99.85\% accuracy. This confirm our theoretical claim for half-twirling case (HT).}
\label{half1}
\end{table}

Now, in the second case, we choose the Hamiltonian is as follows 
\begin{align}
H &= \sigma_{z} \otimes \sigma_{x} + (\xi_{zz} \sigma_{z} \otimes \sigma_{z} + \xi_{0y} \sigma_{0} \otimes \sigma_{y}) .\label{HamilHT2}
\end{align}

Next, we evaluate the contribution of various terms in the Hamiltonian (effective Hamiltonian) both before and after applying HT. We select coherent errors such that three components (ZZ and IY) anti-commute with the original Hamiltonian, and (ZZ and IY anti-commute among themselves).  The simulation parameters for the Hamiltonian in the Eq. (\ref{HamilHT2}) are set to $\xi_{zz} = 0.2$, $\xi_{0y} = 0.2$ and $\tau=0.5$. \textcolor{black}{Table III shows that indeed  the PST fomula does not accurately describe the over-rotation effect when the coherent error terms do not commute among thermselves. We point out that there are cases where the formula
holds even when the terms anti-commute. In other words, the commutation condition is sufficient but not necessary
for the PST formula to apply in HT scenarios.
The fact that HT does exhibit an overrotation is an interesting finding in its own right. By definition, Pauli twirling shows no over-rotation at all. Yet half-twirling, which is composed of half of the Pauli twirling set, does involve an induced overrotation. The conclusion is that the other part of the Pauli twirling set must counteract this induced over-rotation.}

\begin{table}[htbp]
\centering
\begin{tabular}{|c|c|c|c|c|c|}
\hline
 & ZZ & YX & ZX (Numerics) & ZX HT (Theory) & ZX PST (Theory) \\
\hline
No Twirling & 0.2 & 0.2 & 1.0 & - & - \\
\hline
PST & 0 & 0 & 1.02226 & - & 1.02181 \\
\hline
HT & 0 & 0 & 1.0249 & - & - \\
\hline

\end{tabular}
\caption{Comparison of the effective Hamiltonian terms in the evolution operators with and without PST and Half-Twirling (HT) for various coherent error components. While HT and PST completely eliminate the coherent errors in ZZ and YX, the behavior of HT and PST is not the same in this case, as we now have an anti-commuting coherent error. The analytical formula (\ref{eq:=000020main=000020result}) gives the value 1.02181, which matches the numerical result of PST with 99.95\% accuracy, while HT matches the theory prediction with 99.69\% accuracy. This confirms our theoretical claim for the HT case, which requires a commuting coherent error component to behave similarly to PST.}
\label{half2}
\end{table}

\section{Discussion and operational implications}

In practice, when calibrating a gate to generate a rotation of $\theta$,
the drive amplitude (which is proportional to $\tau$) is scanned
until the monitored expectation value matches its ideal values. In
the absence of errors, the scan yields the value $\tau=\theta/2$.
However, in the presence of coherent errors, the scan will lead to
a value of $\tau$ that satisfies $\tau[1+\frac{1-\text{sinc}(2\tau)}{2}(\sum_{\gamma\in\{\gamma^{+}\}}h_{\gamma}^{2})]=\theta/2$.
We emphasize that the calibration process will determine the correct
value of $\tau$ without knowing the values of $h_{\gamma}\LyXZeroWidthSpace$.
As a result, the final accuracy of the gates is not degraded by pseudo-twirling
over-rotation effect studied here. The pseudo-twirling over-rotation
makes the calibration curve slightly non-linear with respect to the
drive amplitude. Consequently, it is not possible to use linearity
to automatically calibrate the gate to a different value of $\tau$.
For instance, to achieve half the rotation, one cannot simply set
$\tau\to\tau/2$.

In this work, we studied a pseudo-twirling induced coherent error
effect (nonlinear over-rotation) arising from the second-order Magnus
expansion. It was explained why this term can be safely ignored in
most cases. Nevertheless, our findings substantiate the validity and
understanding of the pseudo-twirling framework. In particular, by
evaluating the second order Magnus term we learn on the validity regime
of the previous first order Magnus expansion analysis. In this work, we also
point out that, in principle, it is possible to measure the non-linearity
of the actual rotation with respect to the drive amplitude and deduce
the magnitude of the non-commuting coherent errors $\sum_{\gamma\in\{\gamma^{+}\}}h_{\gamma}^{2}$
without carrying out time-consuming process tomography. Finally, the finding that half-twirling also involves induced over-rotation can be of importance for practical purposes. It suggests that calibration with and without HT will yield slightly different results. It is advisable then to carry out the calibration with the HT or PST. This will not only take care of the over-rotation \textcolor{black}{will also} remove spurious coherent errors that interfere in the calibration process.
\section{ACKNOWLEDGMENTS}
Raam Uzdin is grateful for support from the Israel Science Foundation (Grant No. 2556/20). The support of the Israel Innovation Authority is greatly appreciated.

\bibliographystyle{unsrt} 
\bibliography{Refs_PST}

\end{document}